\documentclass[prl,twocolumn,amsfonts,amsmath,tightenlines,preprintnumbers,nofootinbib,superscriptaddress,letterpaper]{revtex4}
\usepackage{graphicx}
\usepackage{bm}

\newcommand{\lsim}{\raisebox{-0.7ex}{$\stackrel{\textstyle <}{\sim}$ }}

\def\si{^1 \hskip -0.03in S _0}
\def\siii{^3 \hskip -0.025in S _1}
\def\diii{^3 \hskip -0.03in D _1}

\begin{document}

\preprint{UNH-12-03}
\preprint{JLAB-THY-12-1509}
\preprint{NT@UW-12-06}
\preprint{NT-LBNL-12-005}
\preprint{UCB-NPAT-12-005}

\begin{figure}[!t]
\vskip -1.1cm
\leftline{
\includegraphics[width=3.0 cm]{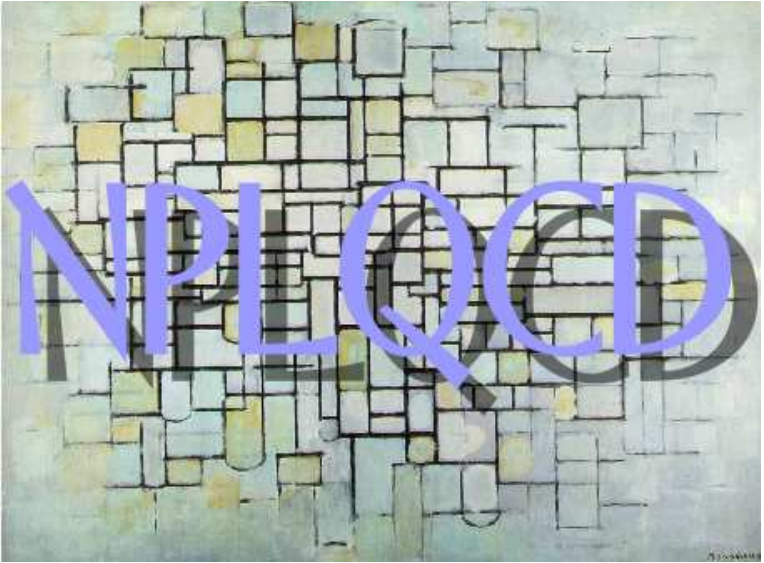}}
\vskip -0.5cm
\end{figure}

\title{Hyperon-Nucleon Interactions 
and the Composition of \\
Dense Nuclear Matter 
from Quantum Chromodynamics}

\author{S.R.~Beane} 
\affiliation{Department of Physics, University
  of New Hampshire, Durham, NH 03824-3568, USA}
\author{E.~Chang}
\affiliation{Dept. d'Estructura i Constituents de la Mat\`eria. 
Institut de Ci\`encies del Cosmos (ICC),
Universitat de Barcelona, Mart\'{\i} Franqu\`es 1, E08028-Spain}
\author{S.D.~Cohen}
\affiliation{Department
  of Physics, University of Washington, Seattle, WA 98195-1560.}
\author{W.~Detmold} 
\affiliation{Department of Physics, College of William and Mary, Williamsburg,
  VA 23187-8795, USA}
\affiliation{Jefferson Laboratory, 12000 Jefferson Avenue, 
Newport News, VA 23606, USA}
\author{H.-W.~Lin}
\affiliation{Department
  of Physics, University of Washington, Seattle, WA 98195-1560.}
\author{T.C.~Luu}
\affiliation{N Division, Lawrence Livermore National Laboratory, Livermore, CA
  94551, USA}
\author{K.~Orginos}
\affiliation{Department of Physics, College of William and Mary, Williamsburg,
  VA 23187-8795, USA}
\affiliation{Jefferson Laboratory, 12000 Jefferson Avenue, 
Newport News, VA 23606, USA}
\author{A.~Parre\~no}
\affiliation{Dept. d'Estructura i Constituents de la Mat\`eria. 
Institut de Ci\`encies del Cosmos (ICC),
Universitat de Barcelona, Mart\'{\i} Franqu\`es 1, E08028-Spain}
\author{M.J.~Savage} 
\affiliation{Department
  of Physics, University of Washington, Seattle, WA 98195-1560.}
\author{A.~Walker-Loud}
\affiliation{Lawrence Berkeley National Laboratory, Berkeley, CA 94720, USA}

\collaboration{ NPLQCD Collaboration }

\date\today

\begin{abstract}
  \noindent 
  The low-energy $n\Sigma^-$ interactions determine, in part, the role
  of the strange quark in dense matter, such as that found in
  astrophysical environments.  The scattering phase shifts for this
  system are obtained from a numerical evaluation of the QCD path
  integral using the technique of lattice QCD.  Our calculations,
  performed at a pion mass of $m_\pi\sim 389~{\rm MeV}$ in two large
  lattice volumes, and at one lattice spacing, are extrapolated to the
  physical pion mass using effective field theory.  The interactions
  determined from QCD are consistent with those extracted from
  hyperon-nucleon experimental data within uncertainties, and
  strengthen theoretical arguments that the strange quark is a crucial
  component of dense nuclear matter.
\end{abstract}

\maketitle


\noindent
The interactions between hyperons and nucleons are important for
understanding the composition of dense nuclear matter.  In
high-density baryonic systems, the large values of the Fermi energies
may make it energetically advantageous for some of the nucleons to
transform into hyperons via the weak interactions, with the increase
in rest mass being more than compensated for by the decrease in
combined Fermi energy of the baryon-lepton system.  This is speculated
to occur in the interior of neutron stars, but a quantitative
understanding of this phenomenon depends on knowledge of the
hyperon-nucleon (YN) interactions in the medium.  In this letter we
use $n\Sigma^-$ scattering phase shifts in the $\si$ and $\siii$
spin-channels calculated with Lattice QCD (LQCD) to quantify the
energy shift of the $\Sigma^-$ hyperon in dense neutron matter, as
might occur in the interior of a neutron star.  Our results strongly
suggest an important role for strangeness in such environments.

Precise nucleon-nucleon (NN) interactions constrained by experiment
and chiral symmetry, together with numerically small but important
three-nucleon interactions, have served as input to refined many-body
techniques for studying the structure of nuclei, such as
Green-function Monte-Carlo~\cite{Pieper:2007ax}, the No-Core Shell
Model \cite{Navratil:2009ut}, or lattice effective field theory 
\cite{Epelbaum:2009pd}, which have led to
remarkably successful calculations of the ground states and excited
states of light nuclei, with atomic number $A<14$. By contrast, the
YN potentials, which are essential for a
first-principles understanding of the hypernuclei and dense matter,
are only very-approximately known. Therefore, gaining a quantitative
understanding of YN interactions --- on a par with knowledge of
the NN interactions --- through experimental and LQCD methods, is a
fundamental goal of nuclear science.

Existing experimental information about the YN interaction comes from
the study of hypernuclei~\cite{hypernuclei-review,Hashimoto:2006aw},
the analysis of associated $\Lambda$-kaon and $\Sigma$-kaon production
in NN collisions near
threshold~\cite{Ba98,Bi98,Se99,Ko04,AB04,GHHS04}, hadronic
atoms~\cite{Batty:1997zp}, and from charge-exchange production of
hyperons in emulsions and pixelated scintillation
devices~\cite{Ahn:2005gb}.  There are only a small set of
cross-section measurements of the YN processes, and not surprisingly,
the extracted scattering parameters are not accurately known. The
potentials developed by the Nijmegen~\cite{nij99,nij06} and
J\"ulich~\cite{HHS89,RHKS96,HM05} groups are just two examples of
phenomenological models based on meson exchange, but the couplings in
such models are not completely determined by the NN interaction and
are instead obtained by a fit to the available YN data.  In
Refs.~\cite{nij99,nij06}, for example, six different models are
constructed, each describing the available YN cross-section data
equally well, but predicting different values for the phase shifts.
Effective field theory (EFT) descriptions have also been
developed~\cite{savage-wise,KDT01,Hammer02,BBPS05,PHM06} and have the
advantage of being model independent.

In the absence of precise experimental measurements, LQCD
calculations can be used to constrain the YN interactions. Several
years ago, the NPLQCD Collaboration performed the first $n_f=2+1$ LQCD
calculations of YN interactions~\cite{Beane:2006gf} (and NN
interactions~\cite{Beane:2006mx}) at unphysical pion masses. Quenched
and dynamical calculations were subsequently performed by the HALQCD
Collaboration~\cite{Nemura:2008sp} and by NPLQCD~\cite{Beane:2009py}.
Recent work by NPLQCD~\cite{Beane:2010hg,Beane:2011xf,Beane:2011iw}
and HALQCD~\cite{Inoue:2010es,Inoue:2011ai} has shown that the $S=-2$
H-dibaryon is bound for pion masses larger than those of nature, and
NPLQCD~\cite{Beane:2011iw} has shown that the same is true for the
$\Xi^-\Xi^-$ with $S=-4$.  In this letter, we use the results of LQCD
calculations to determine leading-order (LO) couplings of the YN EFT
(using Weinberg power counting~\cite{PHM06}) which in turn allow for a
determination of YN interactions at the physical pion mass.

%
In LQCD, L\"uscher's
method~\cite{Hamber:1983vu,Luscher:1986pf,Luscher:1990ux,Beane:2003da}
can be employed to extract two-particle scattering amplitudes below
inelastic thresholds.  For a single scattering
channel, the deviation of the energy eigenvalues of the two-hadron
system in the lattice volume from the sum of the single-hadron masses
is related to the scattering phase shift $\delta(q)$.  The Euclidean
time behavior of LQCD correlation functions of the form
$C_\chi(t)=\langle 0 |\chi(t)\chi^\dagger(0)|0\rangle$, where $\chi$
represents an interpolating operator with the quantum numbers of the
one-particle or two-particle systems under consideration, determines
the ground-state energies of the one-particle and two-particle
systems, $E^{A,B}_1=m_{A,B}$ and $E_2^{(AB)} = \sqrt{q^2 + m_A^2} +
\sqrt{q^2 + m_B^2}$, respectively.  The form of the interpolating
operators and the methodology used for extracting the energy shift are
discussed in detail in Ref.~\cite{Beane:2010em}.  By computing the
masses of the particles and the ground-state energy of the two-particle
system, one obtains the squared momentum $q^2$, which can be either
positive or negative.  For s-wave scattering below inelastic
thresholds, $q^2$ is related to the real part of the inverse
scattering amplitude through the eigenvalue
equation~\cite{Luscher:1986pf} (neglecting phase shifts in $l\ge 4$
partial-waves):
\begin{eqnarray}
  \label{eq:2}
  q \cot \delta(q) 
& = & 
  \frac{1}{\pi L} \lim_{\Lambda\to\infty} \sum_{\bf j}^{|{\bf
      j}|<\Lambda}\frac{1}{|{\bf j}|^2 - q^2 \left(\frac{L}{2\pi}\right)^2}  -4\pi \Lambda\,.
\end{eqnarray}
This relation  enables an LQCD determination of the value of 
the phase shift at the momentum $\sqrt{q^2}$.

Determining the ground-state energy of a system in multiple lattice
volumes allows for bound states to be distinguished from scattering
states.  A bound state corresponds to a pole in the S matrix, and in
the case of a single scattering channel, is signaled by
$\cot\delta(q)\rightarrow +i$ in the large-volume limit.  With
calculations in two or more lattice volumes that both have $q^2<0$ and
$q\cot\delta(q)<0$ it is possible using Eq.~(\ref{eq:2}) to perform an
extrapolation to infinite volume to determine the binding energy of
the bound state $B_{\infty}=\gamma^2/m$, where $\gamma$ is the binding
momentum~\cite{Luscher:1986pf,Luscher:1990ux,Beane:2003da}. The range
of nuclear interactions is determined by the pion mass, and therefore
the use of L\"uscher's method requires that $m_\pi L\gg 1$ to strongly
suppress the contributions that depend exponentially upon the volume,
$e^{-m_\pi L}$~\cite{Sato:2007ms}.  However, corrections of the form
$e^{-\gamma L}$, where $\gamma^{-1}$ is approximately the size of the
bound state, must also be small for the infinite volume extrapolation
to rapidly converge.

Our results are from calculations on two ensembles of $n_f=2+1$
anisotropic clover gauge-field
configurations~\cite{Lin:2008pr,Edwards:2008ja} at a pion mass of
$m_\pi\sim 389$ MeV, a spatial lattice spacing of $b_s\sim
0.123(1)~{\rm fm}$, an anisotropy of $\xi\sim 3.5$, and with spatial
extents of $24$ and $32$ lattice sites, corresponding to spatial
dimensions of $L\sim 3.0$ and $3.9~{\rm fm}$, respectively, and
temporal extents of $128$, and $256$ lattice sites, respectively.  A
detailed analysis demonstrates that the single-baryon masses in these
lattice ensembles are effectively in the infinite-volume
limit~\cite{Beane:2011pc}, and that exponential volume corrections can
be neglected in this work.  L\"uscher's method assumes that the
continuum single-hadron energy-momentum relation is satisfied over the
range of energies used in the eigenvalue equation in
Eq.~(\ref{eq:2}). As discussed in
Refs.~\cite{Beane:2010hg,Beane:2011iw}, the uncertainties in the
energy-momentum relation translate to a $2\%$ uncertainty in the
determination of $q^2$.

We focus on $\si$ and $\siii$ $n\Sigma^-$ interactions, $N\Sigma$ in
the $I=3/2$ channel, and do not consider the $I=1/2$
$N\Sigma$-$N\Lambda$ coupled channels. Calculations in the $I=1/2$
channel are complicated by the proximity in energy of the ground and
first-excited levels in the finite volume. Moreover, while the
$\Sigma$ is more massive than the $\Lambda$, the presence of
$\Lambda$'s in dense matter does not lower the electron Fermi energy.
In the limit of SU(3) flavor symmetry, the $\si$-channels are in
symmetric irreducible representations of ${\bf 8}\otimes {\bf 8} =
{\bf 27}\oplus {\bf 10}\oplus \overline{{\bf 10}} \oplus {\bf 8}
\oplus {\bf 8} \oplus {\bf 1}$, and hence the $n\Sigma^-$ (along with
the $\Xi^-\Xi^-$, $\Sigma^-\Sigma^-$, $nn$, and $\Sigma^-\Xi^-$)
transforms in the ${\bf 27}$.  YN and NN scattering data along with
the leading SU(3) breaking effects, arising from the light-meson and
baryon masses, suggest that all of these channels are attractive at
the physical pion mass, and that $\Xi^-\Xi^-$ and $\Sigma^-\Sigma^-$
are bound~\cite{Stoks:1999bz,Miller:2006jf,Haidenbauer:2009qn}.  By
contrast, the $\siii$-channel of $n\Sigma^-$ scattering transforms in
the ${\bf 10}$ in the limit of SU(3) symmetry, and is therefore
unrelated to NN interactions.  Hence, this channel is quite uncertain,
with disagreements among hadronic models as to whether the interaction is
attractive or repulsive.
\begin{figure}[!t]
  \centering
  \includegraphics[width=0.8\columnwidth]{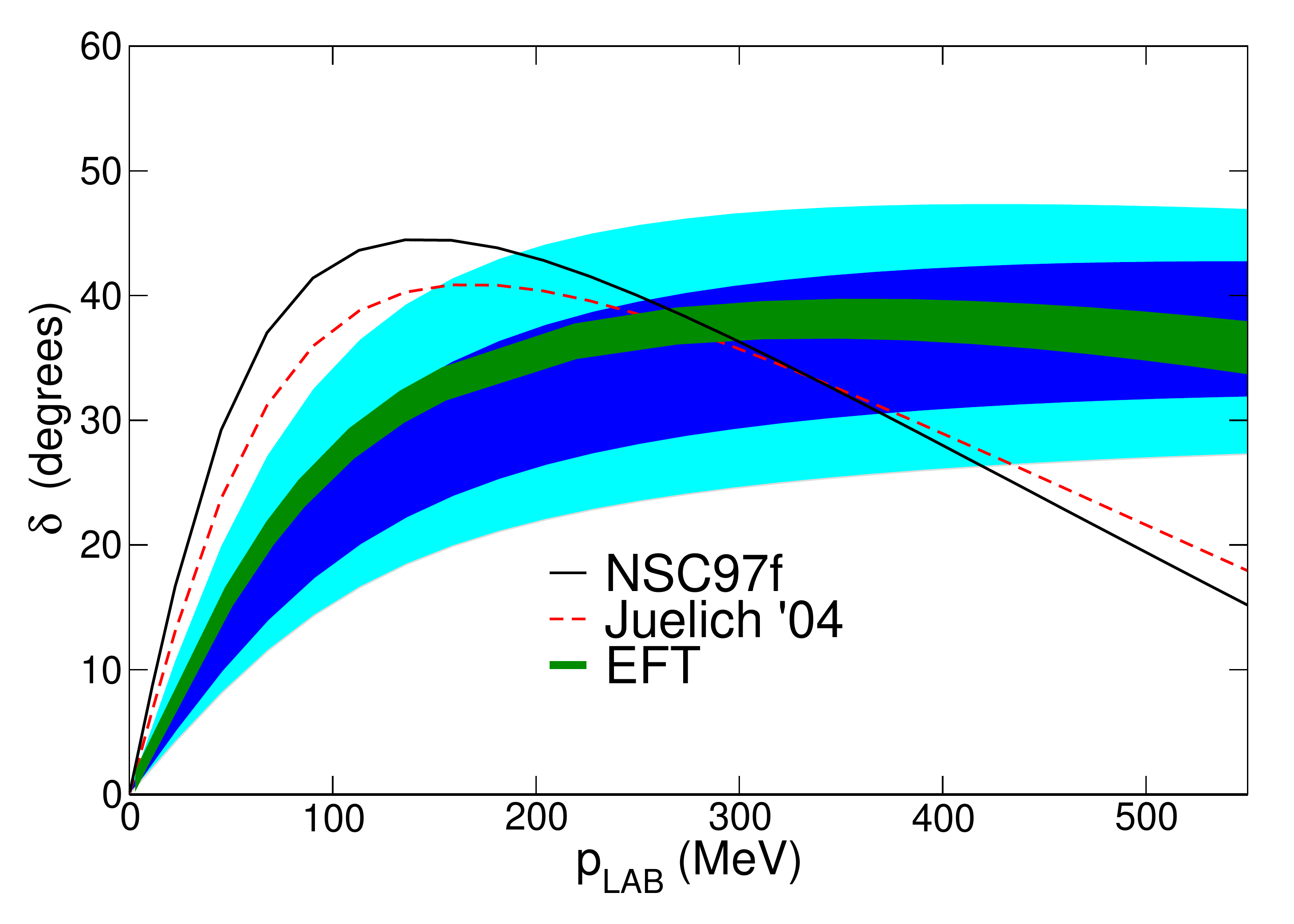}\ \
\vskip -0.15in
   \caption{LQCD-predicted $\si$ $n\Sigma^-$ phase shift versus laboratory momentum at the
   physical pion mass (blue bands), compared with other determinations, as discussed in the text.}
\label{fig:SINGbandd}
\end{figure}

The low-energy $n\Sigma^-$ interactions can be described by an EFT of
nucleons, hyperons and pseudoscalar mesons ($\pi$, K and $\eta$),
constrained by chiral symmetry~\cite{savage-wise,BBPS05,PHM06}.  At
leading order (LO) in the expansion, the $n\Sigma^-$ interaction is
given by one-meson exchange together with a contact operator that
encodes the low-energy effect of short-distance interactions.  As
these contact operators are independent of the light-quark masses, at
LO the quark-mass dependence of the $n\Sigma^-$ interactions is
dictated by the meson masses. Therefore, in each partial wave, a
single lattice datum at a sufficiently low pion mass determines the
coefficient of the contact operator, thereby determining the LO
interaction, including {\it energy-independent} and {\it local}
potentials, wavefunctions and phase shifts, at the physical pion mass.

We find that our LQCD calculations in the $\si$ $n\Sigma^-$ channel
are consistent with the SU(3) symmetry expectations.  At $m_\pi\sim
389$ MeV, using a volume extrapolation as discussed above, we find that
this channel has a bound state, with binding energy $B=25 \pm 9.3 \pm
11~{\rm MeV}$. The quality of the LQCD data in the $\si$ $n\Sigma^-$
channel is comparable to that of its ${\bf 27}$-plet partner
$\Xi^-\Xi^-$, analyzed in detail in Ref.~\cite{Beane:2011iw} (see also
\cite{NPLQCDip}).  In the EFT, the coefficient of the LO contact
operator in this channel is determined by tuning it to reproduce the
LQCD-determined binding energy. We find that this channel becomes
unbound at $m_\pi\lsim 300$ MeV, in agreement with
Ref.~\cite{Haidenbauer:2011za}, which constrained the LO contact
operator using experimental data.
\begin{figure}[!t]
  \centering
  \includegraphics[width=0.8\columnwidth]{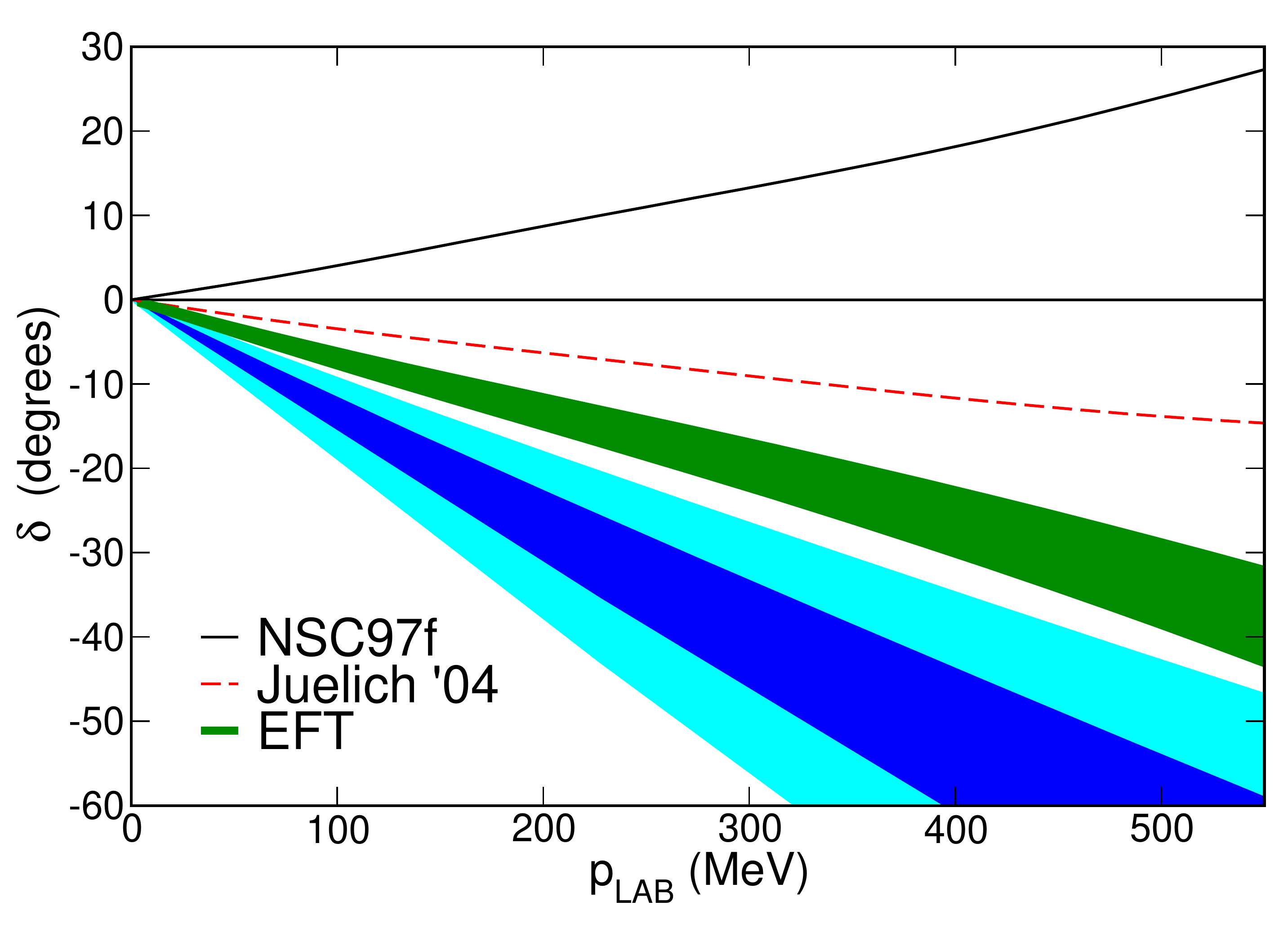}
\vskip -0.15in
   \caption{LQCD-predicted $\siii$ $n\Sigma^-$ phase shift versus laboratory momentum at the
   physical pion mass (blue bands), compared with other determinations, as discussed in the text.}
\label{fig:EMPnsig}
\end{figure}
In Fig.~\ref{fig:SINGbandd}, we show the predicted $\si$ $n\Sigma^-$
phase shift at the physical pion mass --- (dark, light) blue bands
correspond to (statistical, systematic) uncertainties --- and compare
with the EFT constrained by experimental data~\cite{PHM06}, the
Nijmegen NSC97f model~\cite{nij99}, and the J\"ulich '04
model~\cite{HM05}. The systematic uncertainties on our predictions
include those arising from the LQCD calculation (see \cite{NPLQCDip}) as
well as estimates of omitted higher-order effects in the EFT.

The $\siii$-$\diii$ $n\Sigma^-$ coupled channel is found to be highly
repulsive in the s-wave at $m_\pi\sim 389$ MeV, requiring interactions
with a hard repulsive core of extended size.  Such a core, if large
enough, would violate a condition required to use L\"uscher's
relation, namely $R\ll L/2$ where $R$ is the range of the
interaction. We have determined the EFT potential directly by solving
the 3-dimensional Schr\"odinger equation in finite volume to reproduce
the energy levels obtained in the LQCD calculations.  The repulsive
core is found to be large, and formally precludes the use of
L\"uscher's relation, but both methods lead to phase shifts that agree
within uncertainties, indicating that the exponential corrections to
L\"uscher's relation are small.  In Fig.~\ref{fig:EMPnsig}, we show
the predicted $\siii$ $n\Sigma^-$ phase shift at the physical pion
mass.

\begin{figure}[!t]
  \centering  
  \includegraphics[width=0.8\columnwidth]{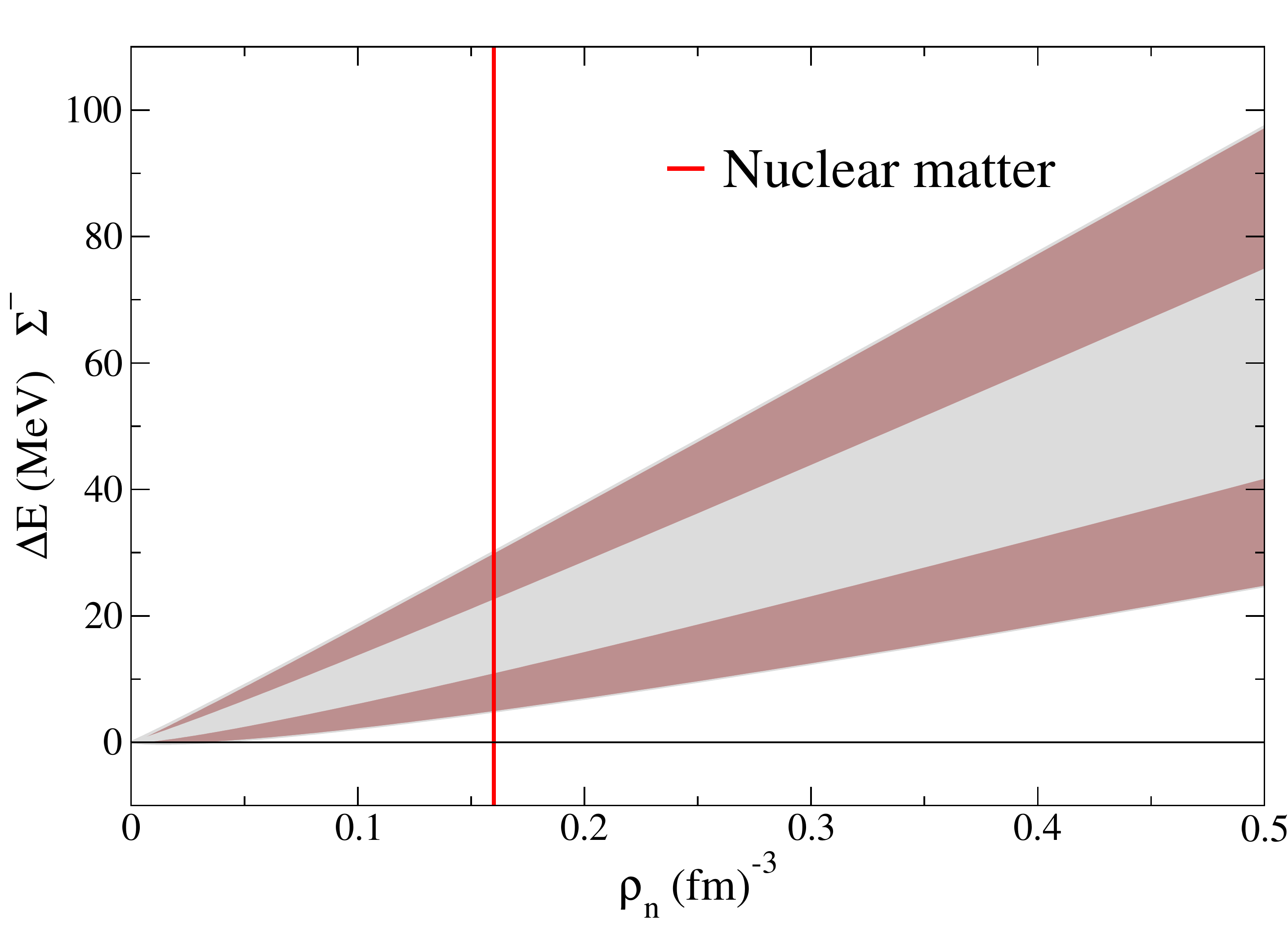}
\vskip -0.15in
  \caption{ The energy shift versus neutron density of a single
    $\Sigma^-$ in a non-interacting Fermi gas of neutrons as
    determined from Fumi's theorem in Eq.~(\protect\ref{eq:fumi}). The
    inner (outer) band encompasses statistical (systematic) uncertainties.}
\label{fig:fumistuff}
\end{figure}

The $n\Sigma^-$ interactions presented here are the crucial
ingredient in calculations that address whether $\Sigma^-$'s
appear in dense neutron matter. As a first step, and in order to understand the
competition between attractive and repulsive components of the $n\Sigma^-$
interaction, we adopt a result due to Fumi for the energy shift due to
a static impurity in a non-interacting Fermi system~\cite{Mahan}:
\begin{equation}
  \label{eq:fumi}
\Delta E \ =  \ -\frac{1}{\pi \mu} \int_0^{k_f} dk\;k\;\Big\lbrack\ \frac{3}{2}\delta_{\siii}(k)\ +\ \frac{1}{2}\delta_{\si}(k)\ \Big\rbrack \ ,
\end{equation}
where $\mu$ is the reduced mass in the $n\Sigma^-$ system.  Using our
LQCD determinations of the phase shifts, and allowing for a $30\%$
theoretical uncertainty, the resulting energy shift and uncertainty
band is shown in Fig.~\ref{fig:fumistuff}.  At neutron number density
$\rho_n\sim0.4~{\rm fm}^{-3}$, which may be found in the interior of
neutron stars, the neutron chemical potential is $\mu_n\sim
M_N+150~{\rm MeV}$ due to neutron-neutron interactions, and the
electron chemical potential, $\mu_{e^-}\sim 200~{\rm
  MeV}$~\cite{Baldo:1999rq}.  Therefore $\mu_n+\mu_{e^-}\sim 1290~{\rm
  MeV}$, and consequently, if $\mu_{\Sigma^-}=M_\Sigma+\Delta E\;\lsim
1290~{\rm MeV}$, that is, $\Delta E\;\lsim 100~{\rm MeV}$, then the
$\Sigma^-$, and hence the strange quark, will play a role in the dense
medium.  We find using Fumi's theorem that $\Delta E= 46\pm
13\pm24~{\rm MeV}$ at $\rho_n=0.4~{\rm fm}^{-3}$.  Corrections due to
correlations among neutrons are difficult to estimate and will require
many-body calculations which are beyond the scope of this
study. Despite this caveat, the results shown in
Fig.~\ref{fig:fumistuff} indicate that the repulsion in the
$n\Sigma^-$ system is inadequate to exclude the presence of
$\Sigma^-$'s in neutron star matter, a conclusion that is consistent
with previous phenomenological modeling (for a review, see
Ref.~\cite{SchaffnerBielich:2010am}).

In this letter, we have presented the first LQCD predictions for
hypernuclear physics, the $\si$ and $\siii$ $n\Sigma^-$ scattering
phase shifts shown in Fig.~\ref{fig:SINGbandd} and
Fig.~\ref{fig:EMPnsig}.  While the LQCD calculations have been
performed at a single lattice spacing, lattice-spacing
artifacts are expected to be smaller than the other systematic
uncertainties. We anticipate systematically refining the analysis
presented in this letter as greater computing resources become
available. The $n\Sigma^-$ interaction is critical in determining the
relevance of hyperons in dense neutron matter, and we have used the
LQCD predictions of the phase shifts to estimate the $\Sigma^-$ energy
shift in the medium.  Our calculation suggests that hyperons are
important degrees of freedom in dense matter, consistent with
expectations based upon the available experimental data and hadronic
modeling.  It is important that more sophisticated many-body
techniques be combined with the interactions determined in this work
to obtain a more precise determination of the energy shift of the
$\Sigma^-$ in medium. This will refine the prediction for the role of
strange quarks in astrophysical environments, and, in particular, will
quantitatively address questions posed by the recent observation of a 
$1.9M_\odot$ neutron star~\cite{Demorest:2010bx}.

\vskip 0.1in

\noindent We thank J.~Haidenbauer, S.~Reddy, K.~Roche and A.~Torok for
valuable conversations, and R. Edwards and B. Jo\'o for help with
QDP++ and Chroma~\cite{Edwards:2004sx}.  We acknowledge computational
support from the USQCD SciDAC project, NERSC and ALCF (Office of
Science of the DOE, DE-AC02-05CH11231 and DE-AC02-06CH11357), the UW
Hyak facility (NSF PHY-09227700), BSC-CNS (Barcelona), LLNL, and
XSEDE, which is supported by NSF grant OCI-1053575.  We acknowledge
support by the NSF grants CAREER PHY-0645570, PHY-0555234, and
CCF-0728915; by DOE grants DE-FG03-97ER4014, DE-AC05-06OR23177,
DE-FG02-04ER41302, OJI DE-SC0001784, DE-FC02-06ER41443,
DE-AC02-05CH11231; by FIS2008-01661 from MEC and FEDER, RTN Flavianet
MRTN-CT-2006-035482; by Jeffress Memorial Trust grant J-968.

%
%


\begin{thebibliography}{99}




\bibitem{Pieper:2007ax} 
  S.~C.~Pieper,
  Riv.\ Nuovo Cim.\  {\bf 31}, 709 (2008).

\bibitem{Navratil:2009ut} 
  P.~Navratil, S.~Quaglioni, I.~Stetcu and B.~R.~Barrett,
  J.\ Phys.\ G G {\bf 36}, 083101 (2009).

\bibitem{Epelbaum:2009pd} 
  E.~Epelbaum, H.~Krebs, D.~Lee and U.~-G.~Meissner,
  Phys.\ Rev.\ Lett.\  {\bf 104}, 142501 (2010).

\bibitem{hypernuclei-review}
A. Gal, E. Hungerford, Nucl. Phys. {\bf A 754} 1-489 (2005).

\bibitem{Hashimoto:2006aw}
  O.~Hashimoto and H.~Tamura,
  Prog.\ Part.\ Nucl.\ Phys.\  {\bf 57}, 564 (2006).

\bibitem{Ba98}
J. Balewski {\it et al.}, Phys. Lett. B {\bf 420}, 211 (1998).

\bibitem{Se99}
S. Sewerin {\it et al.}, Phys. Rev. Lett. {\bf 83}, 682 (1999).

\bibitem{Ko04}
P. Kowina {\it et al.}, Eur. Phys. j. {\bf A22}, 293 (2004).

\bibitem{Bi98}
R. Bilger {\it et al.}, Phys. Lett. B {\bf 420}, 217 (1998).

\bibitem{AB04}
M. Abdel-Bary {\it et al.}, Phys. Lett. B {\bf 595}, 127 (2004).

\bibitem{GHHS04}
A. Gasparyan, J. Haidenbauer, C. Hanhart, J. Speth, Phys. Rev. C {\bf 69}, 
034006 (2004).

\bibitem{Batty:1997zp}
  C.~J.~Batty, E.~Friedman and A.~Gal,
  Phys.\ Rept.\  {\bf 287} (1997) 385.

\bibitem{Ahn:2005gb} 
  J.~K.~Ahn {\it et al.}  [KEK-PS E289 Collaboration],
  Nucl.\ Phys.\ A {\bf 761}, 41 (2005).

\bibitem{nij99}
Th.A.~Rijken, V.G.J.~Stoks and Y.~Yamamoto, Phys. Rev. C {\bf 59}, 21 (1999).

\bibitem{nij06}
Th.A.~Rijken, Y. Yamamoto, Phys. Rev. C {\bf 73} 044008 (2006).

\bibitem{HHS89}
B.~Holzenkamp, K.~Holinde, and J.~Speth, Nucl. Phys. A {\bf 500} (1989) 485.

\bibitem{RHKS96}
A.~Reuber, K.~Holinde, H.-C.~Kim, and J.~Speth, Nucl. Phys. A {\bf 608} 243 (1996).

\bibitem{HM05}
J.~Haidenbauer and U.-G.~ Mei{\ss}ner, Phys. Rev. C {\bf 72} 044005 (2005).

\bibitem{savage-wise}
  M.~J.~Savage and M.~B.~Wise,
  Phys.\ Rev.\ D {\bf 53}, 349 (1996).
  
\bibitem{KDT01}
C.L. Korpa, A.E.L. Dieperink, and R.G.E. Timmermans, Phys. Rev. C {\bf 65}, 015208 (2001).

\bibitem{Hammer02}
H.W. Hammer, Nucl. Phys. A {\bf 705}, 173 (2002). 

\bibitem{BBPS05}
S.R. Beane, P.F. Bedaque, A. Parre\~no, M.J. Savage, Nucl. Phys. A {\bf 747}, 55
(2005).


\bibitem{PHM06}
  H.~Polinder, J.~Haidenbauer and Ulf-G.~Mei\ss ner,
  Nucl.\ Phys.\ A {\bf 779}, 244 (2006)
  [arXiv:nucl-th/0605050].

\bibitem{Beane:2006gf}
  S.~R.~Beane {\it et al.},
                  [NPLQCD],
  Nucl.\ Phys.\  A {\bf 794}, 62 (2007).

\bibitem{Beane:2006mx} 
  S.~R.~Beane, P.~F.~Bedaque, K.~Orginos and M.~J.~Savage,
  Phys.\ Rev.\ Lett.\  {\bf 97}, 012001 (2006)
  [hep-lat/0602010].


\bibitem{Nemura:2008sp}
  H.~Nemura, {\it et al.}, 
  Phys.\ Lett.\  B {\bf 673}, 136 (2009).

\bibitem{Beane:2009py}
  S.~R.~Beane {\it et al.},  [NPLQCD Collaboration],
  Phys.\ Rev.\  D {\bf 81}, 054505 (2010).

\bibitem{Beane:2010hg} 
  S.~R.~Beane {\it et al.}, [NPLQCD Collaboration],
  Phys.\ Rev.\ Lett.\  {\bf 106}, 162001 (2011)
  [arXiv:1012.3812 [hep-lat]].

\bibitem{Beane:2011xf}
  S.~R.~Beane {\it et al.},
  Mod.\ Phys.\ Lett.\  A {\bf 26}, 2587 (2011)
  [arXiv:1103.2821 [hep-lat]].

\bibitem{Beane:2011iw} 
  S.~R.~Beane {\it et al.}  [NPLQCD Collaboration],
  Phys.\ Rev.\ D {\it in press}, 2012
  [arXiv:1109.2889 [hep-lat]].

\bibitem{Inoue:2010es} 
  T.~Inoue {\it et al.}  [HAL QCD Collaboration],
  Phys.\ Rev.\ Lett.\  {\bf 106}, 162002 (2011)
  [arXiv:1012.5928 [hep-lat]].

\bibitem{Inoue:2011ai} 
  T.~Inoue {\it et al.}  [HAL QCD Collaboration],
  arXiv:1112.5926 [hep-lat].

\bibitem{Hamber:1983vu} 
H.~W.~Hamber, {\it et al.}, 
  Nucl.\ Phys.\ B {\bf 225}, 475 (1983).

\bibitem{Luscher:1986pf} M.~L\"uscher,
  Commun.\ Math.\ Phys.\ {\bf 105}, 153 (1986).

\bibitem{Luscher:1990ux} M.~L\"uscher,
  Nucl.\ Phys.\ B {\bf 354}, 531 (1991).

\bibitem{Beane:2003da}
  S.~R.~Beane {\it et al.},
[NPLQCD],
  Phys.\ Lett.\  B {\bf 585}, 106 (2004).


\bibitem{Beane:2010em}
  S.~R.~Beane {\it et al.},
  Prog. \ Part.\ Nucl. \ Phys. {\bf 66}, 1, (2011).

\bibitem{Sato:2007ms}
  I.~Sato and P.~F.~Bedaque,
  Phys.\ Rev.\  D {\bf 76}, 034502 (2007).

 \bibitem{Lin:2008pr}
   H.~W.~Lin {\it et al.}  [HS],
   Phys.\ Rev.\  D {\bf 79}, 034502 (2009).

 \bibitem{Edwards:2008ja}
   R.~G.~Edwards, B.~Joo and H.~W.~Lin,
   Phys.\ Rev.\  D {\bf 78}, 054501 (2008).

\bibitem{Beane:2011pc} 
  S.~R.~Beane {\it et al.}, [NPLQCD],
  Phys.\ Rev.\ D {\bf 84}, 014507 (2011)
  [Phys.\ Rev.\ D {\bf 84}, 039903 (2011)].

\bibitem{Stoks:1999bz}
  V.~G.~J.~Stoks and T.~A.~Rijken,
  Phys.\ Rev.\  C {\bf 59}, 3009 (1999)
  [arXiv:nucl-th/9901028].

\bibitem{Miller:2006jf}
  G.~A.~Miller,
  arXiv:nucl-th/0607006.

\bibitem{Haidenbauer:2009qn}
  J.~Haidenbauer, U.-G.~Mei\ss ner,
  Phys.\ Lett.\  {\bf B684}, 275-280 (2010).
  [arXiv:0907.1395 [nucl-th]].

\bibitem{NPLQCDip} 
  S.~R.~Beane {\it et al.}, [NPLQCD],
{\it in preparation}.

\bibitem{Haidenbauer:2011za} 
  J.~Haidenbauer and U.-G.~Meissner,
  arXiv:1111.4069 [nucl-th] {\it to appear in Nucl. Phys. A}.

\bibitem{Mahan} 
  G.D.~Mahan, {\it Many-Particle Physics}, Plenum Press, NY (1981).

\bibitem{Baldo:1999rq} 
  M.~Baldo, G.~F.~Burgio and H.~J.~Schulze,
  Phys.\ Rev.\ C {\bf 61}, 055801 (2000)
  [nucl-th/9912066].

\bibitem{SchaffnerBielich:2010am} 
  J.~Schaffner-Bielich,
  Nucl.\ Phys.\ A {\bf 835}, 279 (2010)
  [arXiv:1002.1658 [nucl-th]].

\bibitem{Demorest:2010bx} 
  P.~Demorest, T.~Pennucci, S.~Ransom, M.~Roberts and J.~Hessels,
  Nature {\bf 467}, 1081 (2010)
  [arXiv:1010.5788 [astro-ph.HE]].

\bibitem{Edwards:2004sx} R.~G.~Edwards and B.~Joo,
  Nucl.\ Phys.\ Proc.\ Suppl.\ {\bf 140} (2005) 832.





\end{thebibliography}
\end{document}